\documentclass[twocolumn,showpacs,preprintnumbers,amsmath,amssymb]{revtex4}

\usepackage{dcolumn}
\usepackage{bm}

\usepackage{amsmath}
\usepackage{amsfonts}
\usepackage{graphics}
\usepackage{graphicx}
\usepackage{epsfig}
\usepackage{rotating}
 \usepackage[latin1]{inputenc}
 \usepackage{color}

\begin{document}

\preprint{LMU-ASC 38/06}

\title{Coexistence versus extinction in the stochastic cyclic Lotka-Volterra model}

\author{Tobias Reichenbach, Mauro Mobilia, and Erwin Frey}
\affiliation{Arnold Sommerfeld Center for Theoretical Physics and
  Center for NanoScience, Department of Physics,
  Ludwig-Maximilians-Universit\"at M\"unchen, Theresienstrasse 37,
  D-80333 M\"unchen, Germany}


\date{June 8, 2006}
  \begin{abstract}
Cyclic dominance of species has been identified as a potential mechanism
to maintain biodiversity, see e.g. B. Kerr, M. A. Riley, M. W. Feldman and B. J. M. Bohannan [Nature {\bf 418}, 171 (2002)] and B. Kirkup and M. A. Riley [Nature {\bf 428}, 412 (2004)].
Through analytical methods supported by numerical simulations, we address this issue by studying the properties of a paradigmatic  non-spatial three-species stochastic system, namely  the `rock-paper-scissors' or cyclic Lotka-Volterra model. While the
deterministic approach (rate equations) predicts the coexistence of the species
resulting in regular (yet neutrally stable) oscillations of the population densities, we demonstrate that fluctuations arising in the system with a \emph{finite number of agents} drastically alter this picture and are responsible for extinction: After long enough time, two of the three species die out.
As main findings we provide analytic estimates and numerical computation of the extinction probability at a given time.
We also discuss the implications of our results for a broad class of competing population systems.
\end{abstract}

\maketitle

\section{Introduction}

Understanding biodiversity and coevolution is a central challenge in modern evolutionary and theoretical
biology \cite{May,Haken,Neal}. In this context, for some decades much effort has been devoted to mathematically
model dynamics of competing populations through nonlinear, yet deterministic, set of rate equations
like the equations devised by Lotka and Volterra \cite{lotka-1920-42,volterra-1926-31} or many of their variants \cite{May,Haken,Neal}. This
heuristic approach is often termed as population-level description.
As a common feature, these deterministic models  fail to account for stochastic effects (like fluctuations and spatial correlations). However, to gain some more realistic and fundamental understanding
on generic features of population dynamics and mechanisms leading  to biodiversity, it is highly desirable to include internal stochastic noise in the description of agents' kinetics by going beyond the classical deterministic picture.
One of the main reasons is to account for discrete degrees of freedom and finite-size fluctuations \cite{mcadams-1999-15,shnerb-2000-97}. In fact, the deterministic rate equations always (tacitly) assume the presence of infinitely many interacting agents, while in real systems there is  a \emph{large}, yet \emph{finite}, number
of individuals (recently, this issue has been addressed in Refs.~\cite{traulsen-2005-95,traulsen-2006-}).  As a consequence, the dynamics is intrinsically stochastic and the unavoidable finite-size fluctuations may have drastic effects and even completely
invalidate the deterministic predictions.

Interestingly, both {\it in vitro} \cite{kerr-2002-418} and {\it in vivo} \cite{kirkup-2004-428}
experiments  have recently been devoted to experimentally probe
the influence of stochasticity on biodiversity: The authors of Refs.~\cite{kerr-2002-418,kirkup-2004-428} have investigated the mechanism necessary to ensure coexistence in a community of three populations of {\it Escherichia coli} and have numerically modelled the dynamics of their experiments
by the so-called `rock-paper-scissors'  model, well-known in the field of game theory \cite{Hofbauer,worldrps}. This
is a three-species cyclic generalization of the Lotka-Volterra model \cite{frachebourg-1996-54,provata-1999-110,szabo-2001-63}.
As a result, the authors of Ref.~\cite{kerr-2002-418} reported that in a well-mixed (non-spatial) environment (i.e.~when the experiments were carried out in a flask) two species got extinct after some finite time, while coexistence of the populations was never observed. Motivated by these experimental results, in this work we theoretically study the stochastic version of the cyclic Lotka-Volterra model and investigate in detail the effects of finite-size fluctuations on possible population extinction/coexistence.

For our investigation, as suggested by the flask experiment of Ref.~\cite{kerr-2002-418}, the stochastic dynamics of the cyclic Lotka-Volterra model is formulated in the natural language of urn models \cite{Feller} and by adopting the so-called individual-based description \cite{mckane-2004-70}. In the latter, the explicit rules governing the interaction of a \emph{finite number} of individuals with each other are embodied in a master equation. The fluctuations are then specifically accounted for by an appropriate Fokker-Planck equation derived
from the master equation  via a so-called
van Kampen expansion \cite{VanKampen}. This program allows us to quantitatively study the deviations
 of the stochastic dynamics of the cyclic Lotka-Volterra model with respect to the rate equation predictions and to address the question of the extinction probability, the
 computation of which is the main result of this work.
 From a more general
 perspective, we think that our findings have a broad relevance, both theoretical and practical,
 as they shed further  light on how stochastic noise can dramatically affect the properties
 of the numerous nonlinear  systems whose deterministic description, like in the case of the cyclic Lotka-Volterra model,
 predicts the existence of neutrally stable solutions, i.e. cycles in the
 phase portrait \cite{Jordan}.

This paper is organized as follows: The cyclic Lotka-Volterra model is introduced in the next section and its deterministic rate equation treatment is presented. In Section \ref{stoch_appr}, we develop a quantitative analytical approach that accounts for stochasticity, a Fokker-Planck equation is derived from the underlying master equation within a Van Kampen expansion. This allows us to compute the variances of the agents' densities.
We also study the time-dependence properties of the system by carrying out a Fourier analysis from a set of Langevin equations.
Section \ref{sect-ext-prob} is devoted to the computation of the probability of having extinction of two species at a given time, which constitutes the main issue of this work. In the final section, we summarize our findings and present our conclusions.

\section{The cyclic population model: Introduction and  analysis of the rate equations}

\subsection{The model}

\begin{figure}
\begin{center}
\includegraphics[scale=0.45]{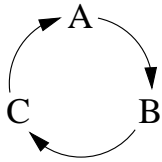}
\caption{Illustration of cyclic dominance of  three states $A$, $B$, and $C$. The latter may correspond to the strategies in a rock-paper-scissors game \cite{Hofbauer}, or to different bacterial species \cite{kerr-2002-418,kirkup-2004-428}.
\label{cycle}}
\end{center}
\end{figure}
The cyclic Lotka Volterra model under consideration here is a system where three states $A$, $B$, and $C$
cyclically dominate each other: $A$ invades $B$, $B$ outperforms $C$, and $C$ in turn dominates over $A$, schematically drawn in Fig.~\ref{cycle}. These three states $A$, $B$, and $C$ allow for various interpretations, ranging from  strategies in the rock-paper-scissors game \cite{worldrps} over tree, fire, ash in forest fire models \cite{clar-1996-8} or chemical reactions \cite{Murray} to different bacterial species \cite{kerr-2002-418,kirkup-2004-428}. In the latter case, a population of poison producing bacteria was brought together with  another being resistant to the poison and a third which is not resistant. As the production of poison as well as the resistance against it have some cost, these species show a cyclic dominance: the poison-producing one invades the non-resistant, which in turn reproduces faster than the resistant one, and  the latter  finally dominates the poison-producing. In a well-mixed environment, like a flask in the experiments \cite{kerr-2002-418,kirkup-2004-428},   eventually only one species survives. The populations are large but finite, and the dynamics of  reproduction and killing events may to a good approximation be viewed as stochastic.

\begin{figure}
\begin{center}
\includegraphics[scale=0.35]{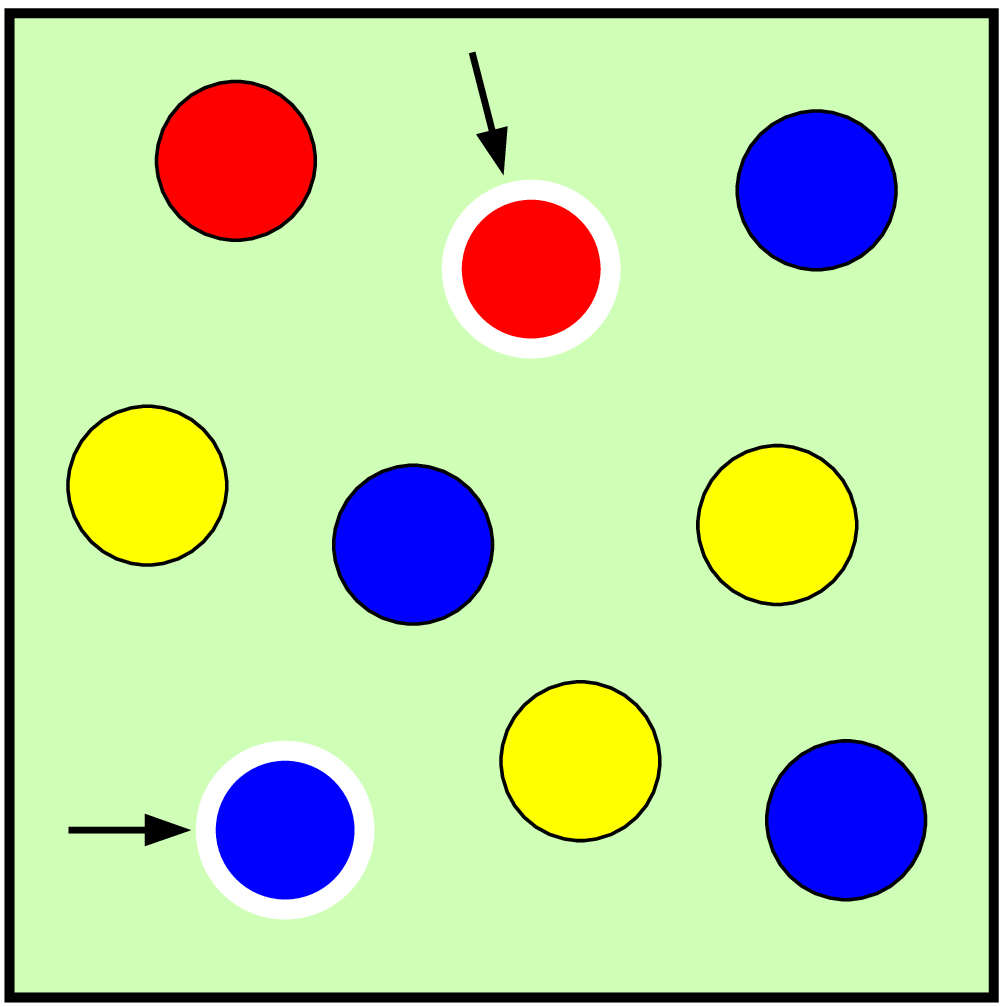}
\hspace{1cm}
\includegraphics[scale=0.35]{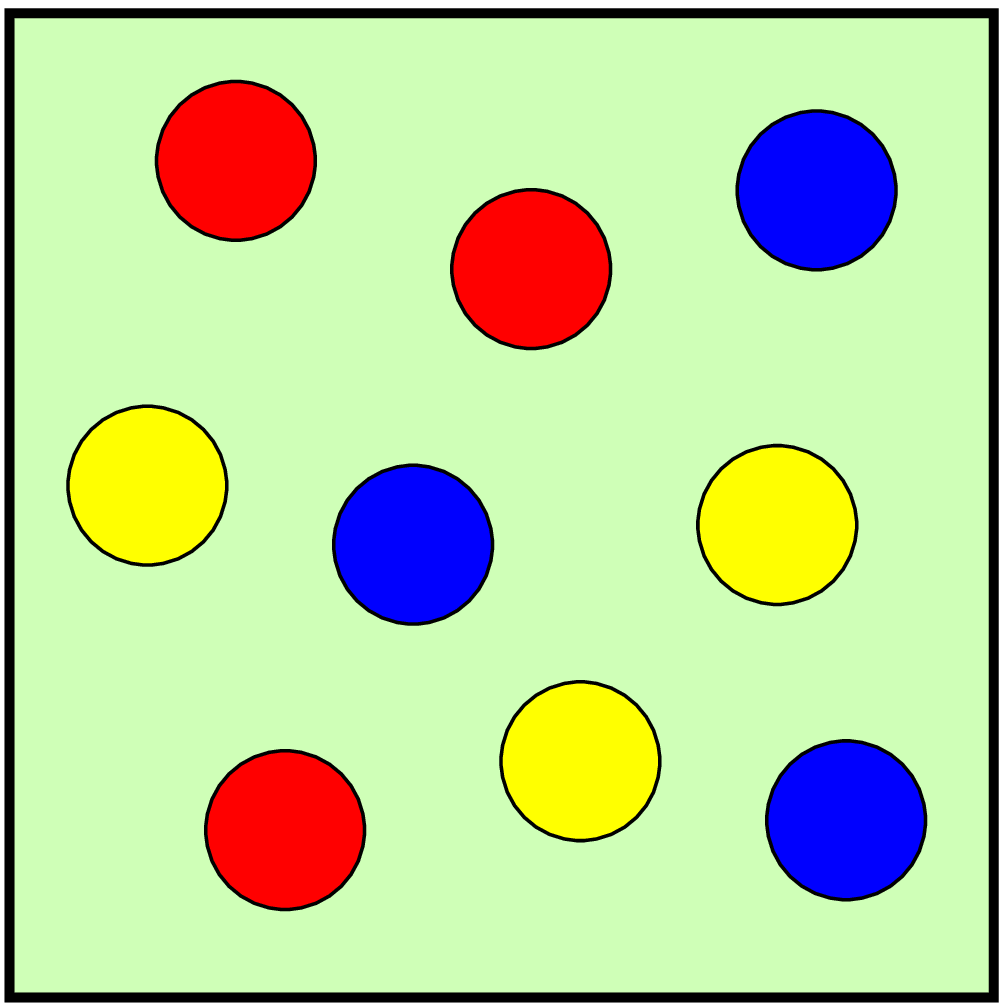}
\caption{(color online) The urn model describes stochastic evolution of well-mixed finite populations. We show three species as yellow or light gray ($A$), red or medium gray ($B$), and blue or dark gray ($C$). At each time step, two random individuals are chosen (indicated by arrows in the left picture) and react (right picture). }
\label{urn}
\end{center}
\end{figure}
Motivated from these biological experiments, we introduce a stochastic version of the cyclic Lotka-Volterra model.
Consider a population of size $N$ which is well mixed, i.e. in the absence of spatial structure. The stochastic dynamics used to describe its evolution, illustrated in Fig.~\ref{urn}, is  referred to as  ``urn model'' \cite{Feller} and closely related to the Moran process \cite{moran-1958-54, nowak-2005-433}. At every time step, two randomly chosen individuals are selected, which may at certain probability  react according to the following scheme:
\begin{align}
A+B&\stackrel{k_C}{\longrightarrow} A+A\cr
B+C&\stackrel{k_A}{\longrightarrow} B+B \label{react} \\
C+A&\stackrel{k_B}{\longrightarrow} C+C\nonumber
\end{align}
with reaction rates $k_A,~k_B$, and $k_C$. We observe the cyclic dominance of the three species. Also, the total number $N$ of individuals is conserved by this dynamics; this
will of course play a role in our further analysis.

We now proceed with the analysis of the deterministic version of the system (\ref{react}).
This will prove insightful for building a stochastic description of the model, which is the scope of
Sec. \ref{stoch_appr}.

\subsection{Deterministic approach\label{det-eq}}

The deterministic rate equations  describe the time evolution of the densities $a(t)$, $b(t)$, and $c(t)$ for the species $A$, $B$, and $C$; they read
 \begin{eqnarray}
\dot{a}&=&a(k_Cb-k_Bc)\nonumber\\
\dot{b}&=&b(k_Ac-k_Ca)\label{RE}   \\
\dot{c}&=&c(k_Ba-k_Ab)\nonumber,
\end{eqnarray}
where the dot stands for the time derivative. These equations describe a well-mixed system, without any spatial correlations,
as naturally implemented in urn models \cite{Feller} or, equivalently, infinite dimensional lattice
systems or complete graphs.
In the following, the Eqs. (\ref{RE}) are discussed  and, from their properties, we  gain
intuition on the effects of stochasticity.

Already from the basic reactions (\ref{react}) we have noticed that the total number of individuals is conserved, which is a property correctly reproduced by the
 the rate equations (\ref{RE}). Setting the total density, meaning the sum of the densities $a$, $b$, and $c$, to unity, we obtain
\begin{equation}
a(t)+b(t)+c(t)=1 \quad
\label{total_dens}
\end{equation}
for all times $t$.
Only two out of the three densities are thus independent, we may view the time evolution of the densities in a two-dimensional phase space.

Eqs.~(\ref{RE}) together with~(\ref{total_dens}) admit three trivial (absorbing) fixed points: $(a^*_{1}, b^*_{1}, c^*_{1})=(1,0,0)$;
$(a^*_{2}, b^*_{2}, c^*_{2})=(0,1,0)$; and $(a^*_{3}, b^*_{3}, c^*_{3})=(0,0,1)$. They denote states where only
one of the three species survived, the other ones died out.  In addition, the rate equations (\ref{RE}) also predict the existence of a fixed point $(a^{*}, b^{*}, c^{*})$ which corresponds to a reactive steady state, associated with the coexistence of all three species:
\begin{align}
a^*&=\frac{k_A}{k_A+k_B+k_C}\cr
b^*&=\frac{k_B}{k_A+k_B+k_C}  \label{FP-c}\\
c^*&=\frac{k_C}{k_A+k_B+k_C} \quad .\nonumber
\end{align}

To determine the nature of this fixed point, we observe that another constant of motion exists for the rate equations (\ref{RE}), namely the quantity
\begin{equation}
a(t)^{k_A}b(t)^{k_B}c(t)^{k_C}\equiv a(0)^{k_A}b(0)^{k_B}c(0)^{k_C}
\label{const2}
\end{equation}
does not evolve in time. In contrast to the total density (\ref{total_dens}), this constant of motion is only conserved by the rate equations but does not stem from the reaction scheme (\ref{react}). Hence, when considering the stochastic version of the cyclic model, the total density remains constant but the expression (\ref{const2}) will no longer be a conserved quantity. The above fixed point (\ref{FP-c}) and constant of motion
(\ref{const2}) have been derived and discussed also within the framework of
game theory, see e.g. Ref.~\cite{Hofbauer}.

\begin{figure}
\begin{center}
\includegraphics[scale=1]{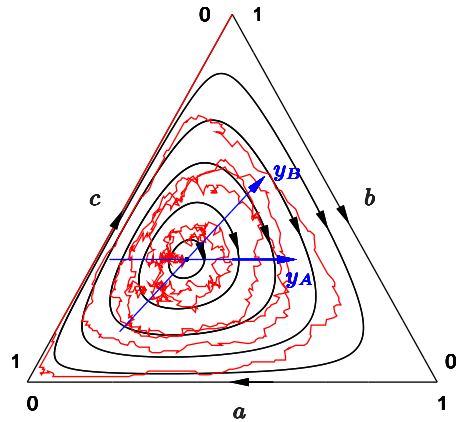}
\caption{(color online) The three-species simplex for reaction rates $k_A=1,~k_B=1.5,~k_C=2$. The rate equations predict cycles, which are shown in black. Their linearization around the reactive fixed point is solved in proper coordinates $y_A,y_B$ (blue or dark gray). The red (or light gray) erratic flow denotes a single trajectory in a finite system ($N=200$), obtained from stochastic simulations. It spirals out from the reactive fixed point, eventually reaching an absorbing state.
\label{simplex_as}}
\end{center}
\end{figure}

In Fig.~\ref{simplex_as}, we depict the ternary phase space~\cite{Rowe} for the densities $a$, $b$, and $c$:  The solutions of the rate equations (\ref{RE}) are shown for different initial conditions and a given set of rates $k_A$, $k_B$, and $k_C$. As the rate equations~(\ref{RE}) are nonlinear in the densities, we cannot solve them analytically, but use numerical methods.
Due to the constant of motion, the solutions  yield cycles around the reactive fixed point (thus corresponding to case 3 in Durrett and Levin's classification~\cite{durrett-1998-53}).
The three trivial steady states, corresponding to saddle points within the linear analysis, are the edges of the simplex. The reactive stationary state, as well as the cycles, are neutrally stable, stemming from the existence of the constant of motion~(\ref{const2}). Especially, the reactive fixed point is a \emph{center fixed point}. The boundary of the simplex denotes states where at least one of the three species died out; as cyclic dominance is lost, states that have reached this boundary will evolve towards one of the edges, making the boundary \emph{absorbing}.

\subsubsection{Linear regime}

The nonlinearity of  Eqs.~(\ref{RE}) induces substantial difficulties in the analytical treatment. However, much can already be inferred from the linearization around the reactive fixed point (\ref{FP-c}), which we will consider in the following.
We therefore introduce the deviations from the reactive fixed point, denoted as $x_A,~x_B,~x_C$:
 \begin{align}
x_A&=a-a^*\cr
x_B&=b-b^* \label{x} \\
x_C&=c-c^*  \qquad .\nonumber
\label{x}
\end{align}
Using conservation of the total density (\ref{total_dens}), we can eliminate $x_C$, and the remaining
linearized equations (\ref{RE}) may be put into the form $\dot{\bf x} = \mathcal{A}{\bf x}$, with the vector ${\bf x}=(x_A,x_B)$ and the matrix
\begin{equation}
\mathcal{A}=\frac{1}{k_A+k_B+k_C}\begin{pmatrix} k_Ak_B & k_A(k_B+k_C)\\ -k_B(k_A+k_C) & -k_Ak_B\end{pmatrix}\quad .
\label{A-lin}
\end{equation}
The reactive fixed point is associated to the eigenvalues $\lambda_{\pm}=\pm i\omega_0$ of $\mathcal{A}$, where $\omega_0$ is given by
\begin{equation}
\omega_0=\sqrt{\frac{k_Ak_Bk_C}{k_A+k_B+k_C}} \quad ;
\label{freq_gen_rates}
\end{equation}
oscillations with this frequency arise in its vicinity. In proper coordinates ${\bf y}=\mathcal{S}{\bf x}$, these oscillations  are harmonic, being the solution of~ ${\bf \dot{y}}=\tilde{\mathcal{A}}{\bf y}$, with  $\tilde{\mathcal{A}}=\mathcal{S}\mathcal{A}\mathcal{S}^{-1}={\footnotesize \begin{pmatrix} 0 & \omega_0 \\ -\omega_0 & 0 \end{pmatrix}}$.  For illustration, we have included these coordinates, in which the solutions take the form of circles around the origin, in Fig.~\ref{simplex_as}. The linear transformation ${\bf x}\rightarrow{\bf y}$ is given by
\begin{equation}
\mathcal{S}=\frac{\sqrt{3}}{2} \begin{pmatrix} \frac{k_A+k_C}{ k_Ak_C}\omega_0 & \frac{1}{k_C} \omega_0\\ 0 & 1\end{pmatrix}\quad .\label{S-matrix}
\end{equation}
The equations $\dot{{\bf y}}=\tilde{\mathcal{A}}{\bf y}$ are easily solved:
\begin{align}
y_A(t)&=y_A(0)\cos(\omega_0 t)+y_B(0)\sin(\omega_0 t)\cr
y_B(t)&=y_B(0)\cos(\omega_0 t)-y_A(0)\sin(\omega_0 t)\quad.
\end{align}
Eventually, we obtain the solutions for the linearized rate equations:
\begin{align}
&a(t)=~a^*+[a(0)-a^*]\cos(\omega_0 t)\cr
&+\omega_0\bigg\{\frac{1}{k_C}[a(0)-a^*]+\frac{k_B+k_C}{k_Bk_C}[b(0)-b^*]\bigg\}\sin(\omega_0 t)\cr
\label{RE-lin-sol}
\end{align}
where $b(t)$ and $c(t)$ follow by cyclic permutations.

\begin{figure}
\begin{center}
\includegraphics[scale=1]{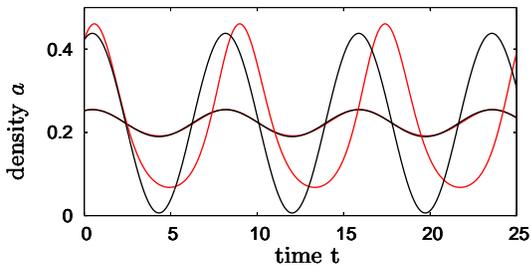}
\caption{ \label{dens_comp} (color online) The deterministic time-evolution of the density $a$ for small and large amplitudes. The prediction (\ref{RE-lin-sol}), shown in black, is compared to the numerical solution (red or gray) of the rate equations (\ref{RE}). For small amplitudes ($a(0)=a^*+0.03,~b(0)=b^*$), both coincide. However, for large amplitudes ($a(0)=a^*+0.2,~b(0)=b^*$), they considerably differ both in amplitude and frequency. We used reaction rates $k_A=1,~k_B=1.5,~k_C=2$.
 }
\end{center}
\end{figure}

To establish the validity of the linear analysis (\ref{A-lin})-(\ref{RE-lin-sol}), we compare the (numerical)
solution of the rate equations (\ref{RE}) with the linear approximations (\ref{RE-lin-sol}). As shown in Fig.~\ref{dens_comp}, when $a(0)-a^*\approx  0.03,~b(0)-b^*=0$, the agreement between the
nonlinear rate equations (\ref{RE}) and the linear approximation (\ref{RE-lin-sol}) is excellent, both curves almost coincide. On the other hand,
the nonlinear terms appearing in Eqs.~(\ref{RE}) become important already when
$a(0)-a^*\approx0.1$ and are responsible for significant discrepancies both in the amplitudes and frequency from the predictions of  Eq.~(\ref{RE-lin-sol}).

\subsubsection{Radius $\mathcal{R}$}

We now aim  at introducing
a measure of  distance $\mathcal{R}$  to the reactive fixed point within the phase portrait.
In the next section, this quantity will help quantify effects of
stochasticity. As it was recently proved useful in a related context ~\cite{traulsen-2005-95},  we aim at taking the structure of cycles predicted by the deterministic equations, see Fig.~\ref{simplex_as}, into account by requiring that the distance should not change on a given cycle. Motivated from the  constant of motion (\ref{const2}), we introduce
\begin{equation}
\mathcal{R}=\mathcal{N}\sqrt{a^{*k_A}b^{*k_B}c^{*k_C}-a^{k_A}b^{k_B}c^{k_C}}\quad,
\label{R-gen}
\end{equation}
with the normalization factor (see below)
\begin{equation}
\mathcal{N}=\sqrt{\frac{3}{2}\frac{k_B(k_A+k_C)}{(k_A+k_B+k_C)^3a^{*k_A}b^{*k_B}c^{*k_C}}} \quad .
\end{equation}
Being conserved by the Eqs.~(\ref{RE}), $\mathcal{R}$  remains constant  on every deterministic cycle. As it vanishes at the reactive fixed point and monotonically grows when departing from it, $\mathcal{R}$ yields a measure of the distance to the latter.

Expanding the radius $\mathcal{R}$ in small deviations ${\bf x}$ from the reactive fixed point results in
\begin{align}
\mathcal{R}^2=&\frac{\mathcal{N}^2}{2}(k_A+k_B+k_C)^2a^{*k_A}b^{*k_B}c^{*k_C}\times\cr
&\times\Big[\Big(\frac{1}{k_A}+\frac{1}{k_C}\Big)x_A^2+\Big(\frac{1}{k_B}+\frac{1}{k_C}\Big)x_B^2+\frac{2}{k_C}x_Ax_B \Big]\cr
&+o({\bf x}^2)\quad .
\label{R-x}
\end{align}
In the variables ${\bf y}=\mathcal{S}{\bf x}$, with our choice for $\mathcal{N}$, it simplifies to
\begin{align}
\mathcal{R}^2=y_A^2+y_B^2+ o({\bf y} ^2)\quad ,
\label{R-y}
\end{align}
corresponding to the radius of the deterministic circles, which emerge in the variables ${\bf y}$.

\subsubsection{Effects of stochasticity: qualitative discussion}

The fact that the number $N$ of particles is  \emph{finite} induces fluctuations  that are not accounted for by Eqs.~(\ref{RE}). In the following, our goal is to understand the importance of fluctuations
and their effects on the deterministic picture (\ref{RE}). We show that, due to the neutrally stable character of the deterministic cycles, fluctuations have drastic consequences. Intuitively,  we expect that in the presence of stochasticity, each trajectory performs a random walk in the phase portrait, interpolating between the deterministic cycles (as will be revealed by considering  $\mathcal{R}$), eventually reaching the boundary of the phase space. There, the cyclic dominance is completely lost, as one of the species gets extinct. Of the two remaining ones, one species is defeating the other, such that the latter soon gets extinct as well, leaving the other one as the only survivor (this corresponds to one of the trivial fixed points, the edges of the ternary phase space). We thus observe the boundary to be absorbing, and presume the system to always end up in one of the absorbing states. A first indication of the actual emergence of this scenario can be inferred from the stochastic trajectory shown in Fig.~\ref{simplex_as}.

\section{Fluctuations in finite populations: a stochastic approach \label{stoch_appr}}

In this section, we set up a stochastic description for the cyclic Lotka-Volterra model in the urn formulation with finite number $N$ of individuals.
Starting from the master equation of the stochastic process,  we obtain a Fokker-Planck equation for the time evolution of the probability $P(a,b,c,t)$ of finding the system in  state $a,b,c$ at time $t$. It allows us to gain a detailed understanding of the stochastic system. In particular, we will  find that, as anticipated at the end of the last section, after long enough time, the system reaches one of the absorbing states. Our main result is the time-dependence of the extinction probability, being the probability that, starting at a situation corresponding to the reactive fixed point, after time $t$ two of the three species have died out. It is obtained through mapping  onto a known first-passage problem. We compare our analytical findings to results from stochastic simulations. For the sake of clarity and without loss of generality, throughout this section, the case of equal reaction rates $k_A=k_B=k_C=1$ is considered. Details on the unequal rates situation are relegated to Appendix~\ref{app_gen_rates}.

\subsection{Stochastic Simulations}

We carried out extensive stochastic simulations to support and corroborate our analytical results. An efficient simulation method originally due to Gillespie \cite{gillespie-1976-22} was implemented for the reactions (\ref{react}). Time and type of the next reaction taking place are determined by random numbers, using the Poisson nature of the individual reactions. For the extinction probability, to unravel the universal time-scaling, system sizes ranging from $N=100$ to $N=1000$ were considered, with sample averages over $10000$ realizations.

\subsection{From the master to Fokker-Planck equation\label{master-f-p}}

Let us start with the master equation of the processes (\ref{react}).  We derive it in the variables $x_A,x_B,x_C$,  which were introduced in (\ref{x}) as the deviations of the densities from the reactive fixed point. Using the conservation of the total density,  $x_A+x_B+x_C=0$,  $x_C$ is eliminated and ${\bf x}=(x_A,x_B)$  kept as independent variables.
The Master equation for the time-evolution of the probability $P({\bf x},t)$ of finding the system in state ${\bf x}$ at time $t$ thus reads
\begin{align}
\partial_t P({\bf x},t)=\sum_{\delta{\bf x}}&\big\{P({\bf x}+\delta{\bf x},t)\mathcal{W}({\bf x}+\delta{\bf x}\rightarrow {\bf x})\cr
&- P({\bf x},t)\mathcal{W}({\bf x}\rightarrow {\bf x}+\delta{\bf x}) \big\}\quad ,
\label{master_eq}
\end{align}
where $\mathcal{W}({\bf x}\rightarrow {\bf x}+\delta{\bf x})$ denotes the transition probability from state ${\bf x}$ to the state ${\bf x}+\delta{\bf x}$ within one time step; summation extends over all possible changes $\delta{\bf x}$. We choose the unit of time so
that, on average, every individual reacts once per time step.\\
According to the Kramers-Moyal expansion of the Master equation to second order in $\delta{\bf x}$, this results in the Fokker-Planck equation
\begin{equation}
\partial_tP({\bf x},t)=-\partial_i[\alpha_i({\bf x})P({\bf x},t)]+\frac{1}{2}\partial_i\partial_j[\mathcal{B}_{ij}({\bf x})P({\bf x},t)] ~.
\label{fokker_planck}
\end{equation}
Here, the indices $i,j$ stand for $A$ and $B$;  in the above equation, the summation convention implies summation over them.
The quantities $\alpha_i$ and $\mathcal{B}_{ij}$ are, according to the Kramers-Moyal expansion:
\begin{align}
\alpha_i({\bf x})=&\sum_{\delta{\bf x}} \delta x_i\mathcal{W}({\bf x}\rightarrow {\bf x}+\delta {\bf x})  \cr
\mathcal{B}_{ij}({\bf x})=&\sum_{\delta {\bf x}}\delta x_i \delta x_j\mathcal{W}({\bf x}\rightarrow {\bf x}+\delta{\bf x}) \quad.
\end{align}
Note that $\mathcal{B}$  is symmetric.
For the sake of clarity, we outline the calculation of $\alpha_A({\bf x})$: The relevant changes $\delta x_A$ in the density $a$ result from the basic reactions (\ref{react}), they are $\delta x_A=1/N$ in the first reaction,  $\delta x_A=0$ in the second and $\delta x_A=-1/N$ in the third. The corresponding rates read  $\mathcal{W}=Nab$ for the first reaction (the prefactor of $N$ enters due to our choice of  time scale, where $N$ reactions occur in one unit of time), and $\mathcal{W}=Nac$ for the third, resulting in $\alpha_A({\bf x})=ab-ac$.
The other quantities are calculated analogously, yielding\\
 \begin{align}
\alpha_A({\bf x})&=ab-ac\cr
\alpha_B({\bf x})&=bc-ab \cr
\mathcal{B}_{AA}({\bf x})&=N^{-1}(ab+ac) \label{alpha_B}\\
\mathcal{B}_{AB}({\bf x})&=\mathcal{B}_{BA}({\bf x})=-N^{-1}ab\cr
\mathcal{B}_{BB}({\bf x})&=N^{-1}(bc+ab)\nonumber
\end{align}
with $a=a^*+x_A,~b=b^*+x_B,~c=c^*-x_A-x_B$.

Van Kampen's linear noise approximation \cite{VanKampen} further simplifies these quantities. In this approach, the
 values $\alpha_i({\bf x})$ are expanded around ${\bf x}=0$ to the first order.
 As they vanish at the reactive fixed point, we obtain:
 \begin{equation}
\alpha_i({\bf x})=\mathcal{A}_{ij}x_j,
\end{equation}
where the matrix elements $\mathcal{A}_{ij}$ are given  by
\begin{equation}
\mathcal{A}_{ij}=\frac{\partial \alpha_i}{\partial x_j}\Big|_{{\bf x}=0} \quad .
\end{equation}
The matrix $\mathcal{A}$, already given in (\ref{A-lin}), embodies the deterministic evolution, while the stochastic noise is encoded in $\mathcal{B}$. To take the fluctuations into account within the Van Kampen expansion, one approximates $\mathcal{B}$  by its values at the reactive fixed point.
Hence, we find:
\begin{align}
\mathcal{A}=\frac{1}{3}\begin{pmatrix} 1 & 2 \\ -2 & -1 \end{pmatrix}\quad, \quad \mathcal{B}=\frac{1}{9N}\begin{pmatrix} 2 & -1 \\ -1 & 2 \end{pmatrix}
\end{align}
and the corresponding Fokker-Planck equation reads
\begin{equation}
\partial_tP({\bf x},t)=-\partial_i[\mathcal{A}_{ij}x_j P({\bf x},t)]+\frac{1}{2}\mathcal{B}_{ij}\partial_i\partial_jP({\bf x},t) \quad .
\label{gen-f-p}
\end{equation}
For further convenience, we now bring Eq.~(\ref{gen-f-p}) into a more suitable form by exploiting
the polar symmetry unveiled by the variables {\bf y}. As for the linearization of
 Eqs.~(\ref{RE}), it is useful to rely on the linear mapping ${\bf x} \to {\bf y}={\cal S}{\bf x}$, with $\mathcal{A}\rightarrow\tilde{\mathcal{A}}=\mathcal{S}\mathcal{A}\mathcal{S}^{-1}$.
Interestingly, it turns out that
this transformation diagonalizes  ${\cal B}$. One indeed finds $\mathcal{B}\rightarrow\tilde{\mathcal{B}}=\mathcal{S}\mathcal{B}\mathcal{S}^T$, with
\begin{equation}
\tilde{\mathcal{B}}=\frac{1}{6N}\begin{pmatrix} 1 & 0\\ 0& 1\end{pmatrix}\quad .
\end{equation}
In the ${\bf y}$ variables, the Fokker-Planck equation (\ref{gen-f-p}) takes the simpler form
\begin{align}
\partial_tP({\bf y},t)=&-\omega_0[y_A\partial_{y_B}-y_B\partial_{y_A}]P({\bf y},t)\cr
&+\frac{1}{12N}[\partial_{y_A}^2+\partial_{y_B}^2]P({\bf y},t)\quad .
\label{f-p-y}
\end{align}
To capture the structure of circles predicted by the deterministic approach,  we introduce polar coordinates:
\begin{align}
y_A&~=r\cos\phi\cr
y_B&~=r\sin\phi\quad.
\end{align}
Note that in the vicinity of the reactive fixed point, $r$ denotes  the distance $\mathcal{R}$, which is now a random variable: $\mathcal{R}=r+o(r)$.
The Fokker-Planck equation (\ref{f-p-y}) eventually turns into
\begin{align}
\partial_t P(r,\phi,t)=&-\omega_0\partial_\phi P(r,\phi,t)\cr
&+\frac{1}{12N}\Big[\frac{1}{r^2}\partial_\phi^2+\frac{1}{r}\partial_r+\partial_r^2\Big]P(r,\phi,t)~ .\cr
\label{pol-f-p}
\end{align}
The first term on the right-hand-side  of this equation describes the system's deterministic evolution, being the motion on circles around the origin at frequency $\omega_0$.
Stochastic effects enter through the second term, which corresponds to isotropic diffusion in two dimensions with diffusion constant $D=1/(12N)$. Note that it vanishes in the limit of infinitely many agents, i.e. when $N\to \infty$.

If we consider a spherically symmetric probability distribution $P(r,\phi,t_0)$ at time $t_0$, i.e. independent of the angle $\phi$ at $t_0$, then this  symmetry is conserved by the dynamics according to Eq.~(\ref{pol-f-p}) and one is left with a radial distribution function $P(r,\phi,t)=P(r,t)$.
%
%
The Fokker-Planck equation (\ref{pol-f-p}) thus further simplifies and reads
\begin{equation}
\partial_t P(r,t)=D\Big[\frac{1}{r}\partial_r+\partial_r^2\Big]P(r,t)\quad .
\label{f_p_r}
\end{equation}
This is the diffusion equation in two dimensions with diffusion constant $D=1/(12N)$, expressed in polar coordinates, for a spherically symmetric probability distribution. This is the
case on which we specifically focus in the following.
Within our approximations around the reactive fixed point, the probability distribution $P(r,t)$ is thus the same as for a system  performing a two-dimensional random walk in the ${\bf y}$ variables.  Intuitively, such behavior is expected and its origin lies in the neutrally stable character of the cycles of the deterministic solution of the rate equations (\ref{RE}). The  cycling around the reactive fixed point does not yield additional effects when considering spherically symmetric probability distributions. Furthermore, due to the existence of the constant of motion (\ref{const2}), the neutral stability does not only hold in the vicinity of the reactive fixed point, but in the whole ternary phase space. We thus expect the distance $\mathcal{R}$ from the reactive fixed point, a random variable, to obey a diffusion-like equation in the whole phase space (and not only around the reactive fixed point). Actually, qualitatively identical behavior is also found in the general case of non-equal reaction rates $k_A$, $k_B$, and $k_C$ (see Appendix \ref{app_gen_rates}), as the constant of motion (\ref{const2}) again guarantees the neutral stability of the deterministic cycles.

\subsection{Fluctuations and frequencies around the reactive fixed point}

In this subsection, we will use the Fokker-Planck equation (\ref{f_p_r}) to investigate the time-dependence of fluctuations around the reactive fixed point. In particular, we are interested in the time evolution of mean deviation from the latter. The average square distance $\langle\mathcal{R}^2(t) \rangle$ will be found to grow linearly
in time, rescaled by the number $N$ of individuals, before saturating.
The frequency spectrum arising from erratic oscillations around the reactive fixed point is of further interest to characterize the stochastic dynamics: We will show
 that the frequency (\ref{freq_gen_rates}) predicted by the deterministic
 approach emerges as a pole in the power spectrum.

\subsubsection{Fluctuations}

\begin{figure}
\begin{center}
\includegraphics[scale=1]{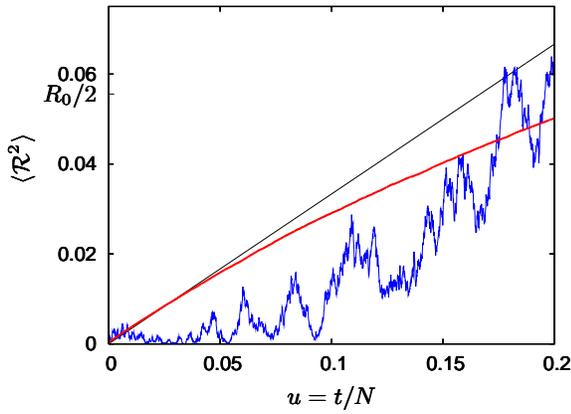}
\caption{ \label{radius_time_evol} (color online) The averaged square radius $\langle \mathcal{R}^2\rangle$ as a function of the rescaled time $u=t/N$. The blue (or dark gray) curve represents a single trajectory, which is seen to fluctuate widely. The linear black line indicates the analytical prediction (\ref{R_time_evol}), the red (or light gray) one corresponds to sample averages over $10^4$ realizations in stochastic simulations. Hereby, we  used a system size of $N=1000$.
 }
\end{center}
\end{figure}
Being interested in the vicinity of the reactive fixed point, in this subsection we ignore  the fact that the boundary of the ternary phase space is absorbing. Then, the solution to the Fokker-Planck equation (\ref{f_p_r}) with  the initial condition $P(r,\phi,t_0=0)=\frac{1}{2\pi r}\delta(r)$, where the prefactor $1/(2\pi r)$ ensures normalization, is simply a Gaussian:
\begin{equation}
P(r,t)=\frac{1}{4\pi D t}\exp\Big(-\frac{r^2}{4Dt}\Big) \quad .
\label{gaussian}
\end{equation}
This result predicts a  broadening of the probability distribution in time, as the average square radius increases linearly with increasing time $t$:
\begin{equation}
\langle r^2(t)\rangle=\frac{1}{3}\frac{t}{N}\quad .
\label{R_time_evol}
\end{equation}
As the  time  scales linearly  with $N$, increasing the system size results in rescaling the time $t$ and the broadening of the probability distribution takes longer. To capture these findings, we introduce the rescaled variable $u\equiv t/N$.

In Fig.~\ref{radius_time_evol}, we compare the time evolution of the squared radius $\langle \mathcal{R}^2(t)\rangle$ obtained from stochastic simulations with the prediction of Eq.~(\ref{R_time_evol}) and find a good agreement in the linear regime around the reactive fixed point,
where $\mathcal{R}=r+o(r)$. In fact, for short times, $\langle \mathcal{R}^2(t)\rangle$ displays a linear time dependence, with systematic deviation from (\ref{R_time_evol})
at longer times. We understand this as being in part due to the linear approximations
used to derive (\ref{R_time_evol}), but also, and more importantly, to the fact that so far we ignored the absorbing character of the boundary. We conclude that the latter invalidates the Gaussian probability distribution for longer times.
This issue, which requires a specific analysis, is the scope of Section
\ref{sect-ext-prob}, where a proper treatment is devised.

\begin{figure}
\begin{center}
\includegraphics[scale=1]{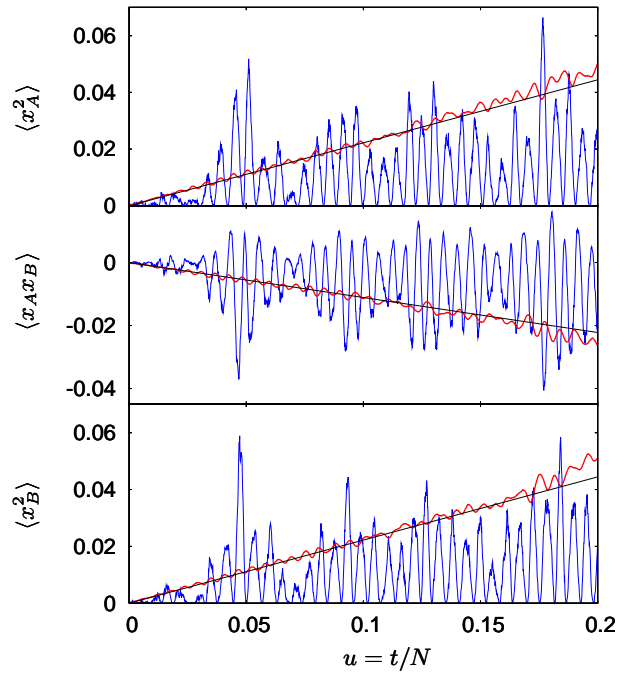}
\caption{ \label{fluct} (color online) Time evolution of the variances of the densities $a$ and $b$ when starting at the reactive fixed point. The blue (or dark gray) curves correspond to a single realization, while the red (or light gray) ones denote averages over 1000 samples. Our results are obtained from stochastic simulations with a system size of $N=1000$. The black line indicates the analytical predictions.
 }
\end{center}
\end{figure}
From the finding $\langle r^2(t)\rangle=2u/3$
together with the spherical symmetry of  (\ref{gaussian}), resulting in $\langle r(t)\phi(t)\rangle=\langle\phi^2(t)\rangle=0$,  we readily obtain the variances of the densities $a$ and $b$:
\begin{align}
\langle x_A^2(t)\rangle=\langle x_B^2(t)\rangle&=\frac{2}{9}u\cr
\langle x_A(t)x_B(t)\rangle&=-\frac{1}{9}u \quad .
\label{fluct_dens}
\end{align}
According to these results, the average square deviations of the densities from the reactive fixed point grow linearly in the rescaled time $u$, thus exhibiting the same behavior as we  already found for the average squared radius
(\ref{R_time_evol}). In Fig.~\ref{fluct}, these findings are compared to stochastic simulations for small times $u$, where the linear growth is indeed recovered.

\begin{figure}
\begin{center}
\includegraphics[scale=1]{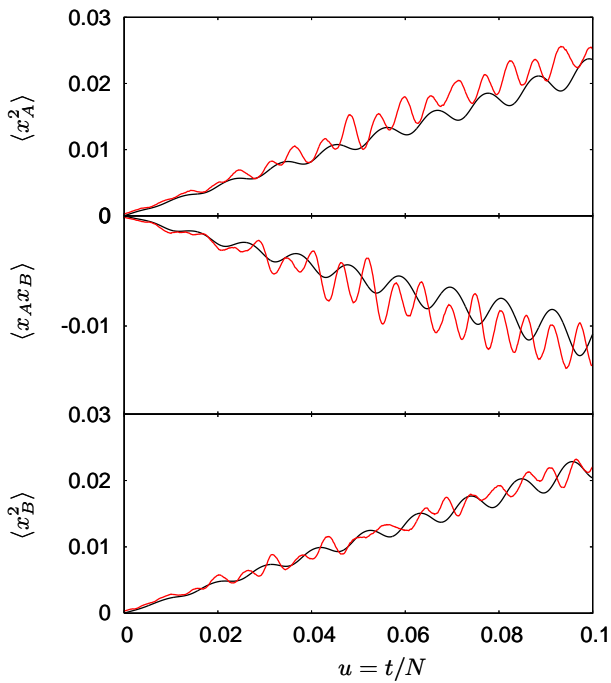}
\caption{  \label{fluct_as} (color online) Variances of the densities $a$ and $b$ when starting at a cycle away from the fixed point. The black lines denote our analytical results, while the red (or gray) ones are obtained by stochastic simulations as averages over 1000 realizations. They are seen to agree for small times, while for larger ones the stochasticity of the system induces erratic oscillations. These results were obtained when starting from $a(t_0=0)=0.37,~b(t_0=0)=1/3$.
 }
\end{center}
\end{figure}
We may consider fluctuations of the densities $a$, $b$, and $c$ not only around the reactive stationary state, as obtained in (\ref{fluct_dens}), but also as the deviations from the deterministic cycles. We consider the latter in the linear approximation around the fixed point, given by Eq.~(\ref{RE-lin-sol}). The fluctuations around them are again described by the Fokker-Planck equation (\ref{fokker_planck}) with $\alpha$ and $\mathcal{B}$
given by Eqs.~(\ref{alpha_B}), but now $x_A$, $x_B$, and $x_C$ are the deviations from the deterministic cycle:
$a(t)=a_0(t)+x_A(t),~b(t)=b_0(t)+x_B(t),~c(t)=c_0(t)-x_A(t)-x_B(t)$, where $a_0$, $b_0$, and $c_0$ obey Eqs.~(\ref{RE}) and characterize the deterministic cycles given by Eq.~(\ref{RE-lin-sol}). Again performing van Kampen's linear noise approximation, we obtain a Fokker-Planck equation of the type (\ref{gen-f-p}), now with the matrices
\begin{align}
\mathcal{A}&=\begin{pmatrix} 2a_0+2b_0-1 & 2a_0 \\ -2b_0 & -2a_0-2b_0+1 \end{pmatrix}\cr
\mathcal{B}&=\frac{1}{N}\begin{pmatrix} a_0(1-a_0) & -a_0b_0 \\ -a_0b_0 & b_0(1-b_0) \end{pmatrix}\quad .
\end{align}
Note that the entries of the above matrices
now depend on time $t$ via $a_0(t)$ and $b_0(t)$.
The Fokker-Planck equation (\ref{gen-f-p}) yields equations for the time evolution of the fluctuations:
\begin{equation}
\partial_t\langle x_ix_j \rangle=\mathcal{A}_{ik}\langle x_k x_j\rangle+\mathcal{A}_{jk}\langle x_k x_i\rangle+\mathcal{B}_{ij} \quad .
\end{equation}
They may be solved numerically for  $\langle x_A^2(t)\rangle$, $\langle x_A(t)x_B(t)\rangle$, and $\langle x_B(t)^2\rangle$, yielding growing oscillations. In Fig.~\ref{fluct_as} we compare these findings to stochastic simulations. We observe a good agreement for small rescaled times $u=t/N$, while the oscillations of the stochastic results become more irregular at longer times. \\

\subsubsection{Frequencies}

Recently, it has been shown that the frequency predicted by the deterministic rate equations
appears in the stochastic system  due to a ``resonance mechanism'' \cite{mckane-2005-94}. Internal noise present in the system covers all frequencies and induces excitations; the largest  occurring for the ``resonant'' frequency predicted by the rate equations.
Here, following the same lines, we address the issue of the characteristic frequency in the stochastic cyclic Lotka-Volterra model.

\begin{figure}
\begin{center}
\includegraphics[scale=1]{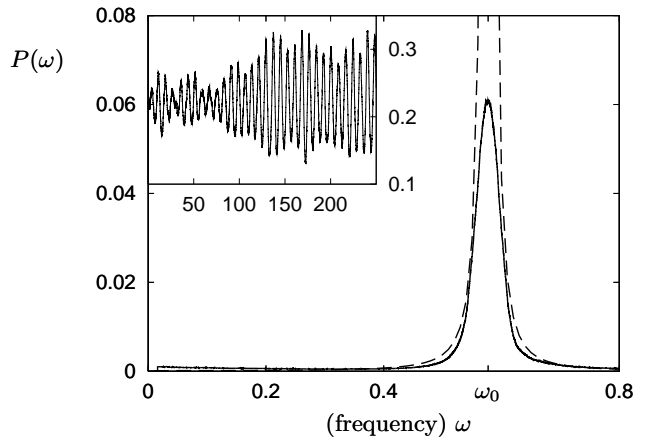}
\caption{The power spectrum for densities in the vicinity of the reactive fixed point. The dashed line indicates our analytical result, which has a pole at $\omega_0$. It agrees  with stochastic simulations (solid). The inset shows the erratic oscillations of the density of one of the species for one realization. The system size considered is $N=10000$.
\label{urn_power_spectrum}}
\end{center}
\end{figure}
In the vicinity of the reactive fixed point, the deterministic rate equations (\ref{RE}) predict density oscillations with frequency  $\omega_0$ given in Eq.~(\ref{freq_gen_rates}). For the stochastic model,  we now show that a spectrum of frequencies centered  around this value $\omega_0$ arises. The most convenient way to compute this power spectrum $P(\omega)\equiv\langle|\tilde{{\bf x}}(\omega)|^2\rangle$ from the Fokker-Planck equation (\ref{gen-f-p}) is through the set of equivalent Langevin equations \cite{Gardiner}:
\begin{align}
\dot{x}_i=\mathcal{A}_{ij}x_j+\xi_i  \quad,
\label{langevin}
\end{align}
with the white noise covariance matrix $\mathcal{B}$:
$\langle \xi_i(t) \xi_j(t')\rangle=\mathcal{B}_{ij}\delta(t-t')$. From the Fourier transform of Eq.~(\ref{langevin}), it follows that
\begin{align}
P(\omega)&\equiv\langle|\tilde{{\bf x}}(\omega)|^2\rangle=\text{Tr}\big[(\mathcal{A}+i\omega)^{-1}\mathcal{B}(\mathcal{A}-i\omega)^{-T}\big]\cr
&=\frac{4}{3N}\frac{1+3\omega^2}{(1-3\omega^2)^2} \quad .
\end{align}
The power spectrum has a pole at the characteristic frequency already predicted from the rate equations (\ref{RE}), $\omega_0=1/\sqrt{3}$.  For increasing system size $N$, the power spectrum displays a sharper alignment with this value.
In Fig.~\ref{urn_power_spectrum} we compare our results to stochastic simulations and find an excellent agreement, except for the pole, where the power spectrum in finite systems obviously has a finite value.
These results were obtained in the vicinity of the reactive fixed point, where the linear analysis (\ref{RE-lin-sol})
applies. As already found in the deterministic description (see Fig.~\ref{dens_comp}), when departing from the center fixed point, nonlinearities will alter the characteristic frequency.

\subsection{The extinction probability\label{sect-ext-prob}}

So far, within the stochastic formulation, the fluctuations around the reactive steady state were found to follow a Gaussian distribution, linearly broadening in time. However, when approaching the absorbing boundary,
the latter alters this behavior, see Fig.~\ref{radius_time_evol}. In the following, we will incorporate
this effect in our quantitative description. It plays an essential role when discussing the extinction probability, which is the scope of this subsection.

The probability that one or more species die out in the course of time is of special interest within population dynamics from a biological viewpoint. When considering meta-populations
formed of local patches, such questions were e.g. raised in Ref.  \cite{alonso-2002-64}. Here, we consider the extinction probability $P_{\text{ext}}(t)$ that, starting at the reactive fixed point, after time $t$ two of the three species have died out.  In our formulation, this corresponds to the probability that after time $t$ the system has reached the absorbing boundary of the ternary phase space, depicted in Fig.~\ref{simplex_as}. Considering states far from the reactive fixed point, we now have to take into account the absorbing nature of the boundary. This feature is incorporated in our approach by discarding the states having reached the boundary, so  that  a vanishing density of states occurs there. As the normalization is lost, we do no longer deal with  probability distributions: In fact, discarding states  at the boundary implies a time decay of the integrated density of states [which, for commodity, we still refer to as $P(r,t)$].
Ignoring the nonlinearities, the problem now takes the form of solving
the Fokker-Planck equation (\ref{f_p_r}) for an initial condition 
$P(r,t_0=0)=\frac{1}{2\pi r}\delta(r)$, and with the requirement that 
$P(r,t)$ has to vanish at the boundary. 
Hereby, the triangular shaped absorbing boundary is 
regarded as the outermost (degenerate) cycle of the deterministic 
solutions. As the linearization in the $y$ variables 
around the reactive fixed point maps the cycles onto circles 
(see above), the triangular boundary is mapped to a sphere as well. 
Although the linearization scheme on which these mappings rely 
is inaccurate in the vicinity of the boundary, it is possible 
to incorporate nonlinear effects in a simple and pragmatic 
manner, as shown below. However, first let us consider the 
linearized problem, being a first-passage to a sphere of radius 
$R$, which is, e.g., treated in Chapter 6 of Ref. \cite{Redner}.
 The solution
is known to be a combination of modified Bessel functions of the first and second kind. Actually, the  Laplace transform of the density of states reads
\begin{widetext}
\begin{align}
\text{L. T.}\{P(r,t) \}=\int dt e^{-st}P(r,t)
=\frac{1}{2\pi D}\frac{I_0\big(R\sqrt{s/D}\big)K_0\big(r\sqrt{s/D}\big)-I_0\big(r\sqrt{s/D}\big)K_0\big(R\sqrt{s/D}\big)}{I_0\big(R\sqrt{s/D}\big)}\cr
\end{align}
\end{widetext}
where $D=1/(12N)$ is the diffusion constant; $I_0$ and $K_0$ denote the Bessel function of the first, resp. second, kind and of order zero; and $R$ is the radius of the
absorbing sphere.
It is normalized  at the initial time $t_0=0$; however, for later times $t$, the total number of states $\int_0^R dr\int_0^{2\pi}d\phi~ rP(r,t)$ will decay in time, as states are absorbed at the boundary.

\begin{figure}
\begin{center}
\includegraphics[scale=1]{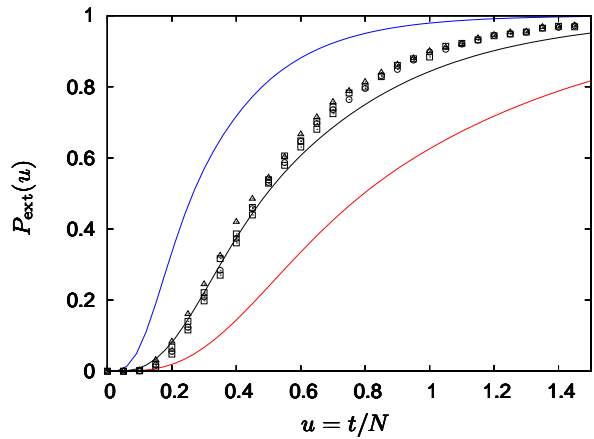}
\caption{ \label{urn_transition} (color online)
The extinction probability  when starting at the reactive fixed point, depending on the rescaled time $u=t/N$. Stochastic simulations for different system sizes ($N=100$: triangles; $N=200$: boxes; $N=500$: circles) are compared to the analytical prediction (\ref{ext_prob_u}). The left (blue or dark gray) is obtained from $R_0=1/3$,  the right one (red or light gray) from $R_1=1/\sqrt{3}$, and the middle (black) corresponds to the average of both: $R_2=\frac{1}{2}(R_0+R_1)$. }
\end{center}
\end{figure}

Equipped with this result, we are now in a position to calculate the extinction probability $P_\text{ext}(t)$. It can be found by considering the probability current $J(\tau)$ at the absorbing boundary \cite{Redner}, namely:
\begin{equation}
J(\tau)=D\int_\text{Sphere}d\Omega~ \frac{\partial P(r,t)}{\partial r}\Big|_{r=R}\quad,
\end{equation}
whose Laplace transform is given by
\begin{equation}
\text{L.T.}(\{J(\tau)\})=\frac{1}{I_0\big(R\sqrt{s/D}\big)}.
\end{equation}
The extinction probability at time $t$  is obtained from $J(\tau)$ by integrating over time $\tau$ until time $t$ \cite{Redner}: $P_\text{ext}(t)=\int_0^td\tau J(\tau)$.
Therefore, the Laplace transform of $P_\text{ext}(t)$ reads
$\text{L.T.}\{P_\text{ext}(t)\}=1/(sI_0(R\sqrt{s/D}))
$. Again, we notice that we can write this equation in a form that depends on the time $t$ only via the rescaled time $u=t/N$:
\begin{equation}
\text{L.T.}\{P_\text{ext}(u)\}=\frac{1}{sI_0(R\sqrt{s/12})}\qquad.
\label{ext_prob_u}
\end{equation}
Hence, for different system sizes $N$, one obtains the same extinction probability
$P_\text{ext}(u)$ provided one considers the same value for the scaling variable $t/N= constant$ (see Fig.~\ref{urn_transition}).

 We cannot solve the inverse Laplace transform appearing in Eq.~(\ref{ext_prob_u}), but $I_0(z)$ might be expanded according to \cite{Abramowitz} as
\begin{equation}
I_0(z)=1+\frac{\frac{1}{4}z^2}{(1!)^2}+\frac{(\frac{1}{4}z^2)^2}{(2!)^2}+\frac{(\frac{1}{4}z^2)^3}{(3!)^2}+...~.
\label{exp_I0}
\end{equation}
Considering the first three terms in this expansion, i.e contributions up to $z^4$,
 yields the approximate result
\begin{equation}
P_\text{ext}(t)=1-\Big(1+\frac{2}{3}\frac{t}{R^2N}\Big) \exp\Big[-\frac{2}{3}\frac{t}{R^2N}\Big]      \quad .
\label{approx_ext_prob}
\end{equation}
Numerically, we have included higher terms. The results for contributions up to $z^{10}$ in (\ref{exp_I0}) are shown in Fig.~\ref{urn_transition}. An estimate of the effective distance $R$ to the absorbing boundary is determined by plugging either $a=0$, or $b=0$, or $c=0$  into Eq.~(\ref{R-gen}), which yields $R_0=1/3$. From Fig.~\ref{urn_transition}, we observe
the  extinction probability to be overestimated by (\ref{approx_ext_prob}) with $R=R_0$. This stems from nonlinearities altering the analysis having led to (\ref{approx_ext_prob}). However, adjusting $R$ on physical grounds, we are able to capture these effects.
Another estimate of $P_\text{ext}$ is obtained by considering the expression Eq.~(\ref{R-x}), which arises from a linear analysis. As the extinction of two species requires $x_A=x_B=1/3$,
 in the expression Eq.~(\ref{R-x}) this leads to the estimate $R_1=1/\sqrt{3}$ for $R$. The comparison with
Fig.~\ref{urn_transition}, shows that (\ref{approx_ext_prob}) together with $R=R_1$
is a lower bound of $P_\text{ext}$. A simple attempt to interpolate between the above estimates is to consider the mean value of these radii, $R_2\equiv\frac{1}{2}(R_0+R_1)$, which happens to yield an excellent agreement with results from stochastic simulations, see Fig.~\ref{urn_transition}.
For the latter, we have considered systems  of $N=100,~ 200$, and $N=500$ individuals. Rescaling the time according to $t/N$, they are seen to collapse on a universal curve. This is well described by Eq.~(\ref{approx_ext_prob}) with $R_2$ as the radius of the absorbing boundary.

To conclude, we  consider the mean time $t_\text{abs}$ that it takes until one species becomes extinct. From \cite{Redner} we find for the rescaled mean absorption time
\begin{equation}
u_\text{abs}=t_\text{abs}/N=3R^2\quad,
\end{equation}
which for our best estimate $R=R_2$ yields a value of $u_\text{abs}\approx 0.62$.

\section{Conclusion}

Motivated by recent {\it in vitro} and {\it in vivo} experiments
aimed at identifying mechanisms responsible for biodiversity
in populations of {\it Escherichia coli} \cite{kerr-2002-418,kirkup-2004-428}, we
have considered the stochastic version of the `rock-paper-scissors', or three-species cyclic Lotka-Volterra, system
within an urn model formulation. This approach allowed us to quantitatively study the effect of finite-size fluctuations in a system with a large, yet \emph{finite}, number $N$ of agents.

While the classical rate equations of the cyclic Lotka-Volterra model predict the existence of one (neutrally stable) center fixed point, associated with the \emph{coexistence} of all the species, this picture is \emph{dramatically}
invalidated by the fluctuations which unavoidably appear in a finite system. The latter were taken into account by a Fokker-Planck equation derived from the underlying master equation through a Van Kampen expansion.

Within this scheme, we were able to show that the variances of the densities of individuals grow in time (first linearly)
until extinction of two of the species occurs. In this context, we have investigated the probability
for such extinction to occur at a given time $t$. As a main result of this work, we have shown that this extinction probability  is a function of the scaling variable $t/N$. Exploiting polar symmetries displayed
by the deterministic trajectories in the phase portrait and using a mapping onto a classical first-passage problem, we were able to provide analytic estimates (upper and lower  bounds) of the extinction probability, which have been successfully compared to numerical computation.

From our results, it turns out that the classical rate equation predictions apply to the urn model with a finite number of agents only for short enough time, i.e. in the regime $t\ll N$. As time increases, the probability of extinction
grows, asymptotically reaching $1$ for $t\gg N$,  so that, for finite $N$, fluctuations are \emph{always} responsible for extinction
and thus dramatically jeopardize the possibility of coexistence and biodiversity.
Interestingly, these findings are in qualitative agreement with those (both experimental and numerical)
reported in Ref.~\cite{kerr-2002-418}, where it was found that in a well mixed environment (as in the urn model considered here) two species get extinct.

While this work has specifically focused on the stochastic cyclic Lotka-Volterra model, the addressed issues are generic.
Indeed, we think that our results and technical approach, here illustrated by considering
the case of a paradigmatic model, might actually shed further light on the role of fluctuations
and the validity of the rate equations in a whole class of stochastic systems.
In fact, while one might believe that fluctuations in an urn model should always vanish in the thermodynamic limit, we have shown that this issue should be dealt with due care: This is true for systems where the rate equations predict the existence of an asymptotically
stable fixed point, which is always reached by the stochastic dynamics \cite{mckane-2005-94,ben-naim-2004-69}.
In  contrast, in systems where the deterministic (rate equation) description
 predicts the existence of (neutrally stable) center fixed points, such as the cyclic Lotka-Volterra model, fluctuations have dramatic consequences and hinder biodiversity by being responsible (at long, yet finite, time) for extinction of species \cite{kerr-2002-418}.
In this case, instead of a deterministic oscillatory behavior around the linearly (neutrally) stable fixed point, the stochastic dynamics always drives the system toward one of its absorbing states.
 Thus, the absorbing fixed points, predicted to be linearly unstable within the rate equation theory, actually turn out to be the \emph{only stable fixed points} available at long time.

\section{Acknowledgments}
We would like to thank U. C. T\"auber and P. L. Krapivsky for helpful discussions, as well as A. Traulsen and J. C. Claussen for having made manuscript \cite{traulsen-2006-} available to us prior to publication.
M.M. gratefully acknowledges the support of the German Alexander
von Humboldt Foundation through the Fellowship No. IV-SCZ/1119205~STP.

\begin{appendix}

\section{Unequal reaction rates reconsidered\label{app_gen_rates}}

In Section \ref{stoch_appr}, when considering the stochastic approach to the cyclic Lotka-Volterra model, for the sake of clarity
we have specifically turned to the situation of equal reaction rates, $k_A=k_B=k_C=1$. In this Appendix, we want to provide some details
on the general case with unequal rates. While the mathematical treatment becomes more involved, we will argue that the qualitative general
situation still follows along the same lines as the (simpler) case that we have discussed in detail in Sec.~\ref{stoch_appr}.

The derivation of the Fokker-Planck equation (\ref{gen-f-p}) is straightforward, following the lines of Subsection \ref{master-f-p}. The matrix $\mathcal{A}$ remains unchanged and is given in Eq.~(\ref{A-lin}), for $\mathcal{B}$ we now obtain:
\begin{align} \mathcal{B}=\frac{1}{N}\frac{\omega_0^2}{k_A+k_B+k_C}\begin{pmatrix} 2 & -1 \\ -1 & 2 \end{pmatrix}\quad .
\end{align}
The corresponding Fokker-Planck equation reads
\begin{equation}
\partial_tP({\bf x},t)=-\partial_i[\mathcal{A}_{ij}x_j P({\bf x},t)]+\frac{1}{2}\mathcal{B}_{ij}\partial_i\partial_jP({\bf x},t) \quad .
\end{equation}
Again, we aim at benefitting from the cyclic structure of the deterministic solutions, and perform a variable transformation to ${\bf y}=\mathcal{S}{\bf x}$, with $\mathcal{S}$ given in Eq.~(\ref{S-matrix}).
As already found in Subsection \ref{det-eq}, $\mathcal{A}$ turns into $\tilde{\mathcal{A}}={\footnotesize \begin{pmatrix} 0 & \omega_0 \\ -\omega_0 & 0 \end{pmatrix}}$, such that in the ${\bf y}$ variables  the deterministic solutions correspond to circles around the origin. For the stochastic part, entering via $\mathcal{B}$, a technical difficulty arises. It is transformed into
\begin{equation}
\tilde{\mathcal{B}}=\frac{3}{2}\frac{1}{N}\frac{\omega_0^2}{k_A+k_B+k_C}\begin{pmatrix} \frac{k_A^2+k_Ak_C+k_C^2}{k_A^2k_C^2}\omega_0^2  & \frac{k_A-k_C}{2k_Ak_C}\omega_0\\ \frac{k_A-k_C}{2k_Ak_C}\omega_0 & 1\end{pmatrix}\quad ,
\end{equation}
being no longer proportional to the unit matrix as in the case of equal reaction rates. We can do slightly better by using an additional rotation, ${\bf z}={\footnotesize \begin{pmatrix} \cos\theta&\sin\theta  \\ -\sin\theta & \cos\theta \end{pmatrix}}{\bf y}$, with rotation angle
\begin{equation}
\tan(2\theta)=\frac{3(k_C-k_A)\sqrt{k_Ak_Bk_C(k_A+k_B+k_C)}}{k_A^2(k_C-k_B)+k_C^2(k_A-k_B)}\quad.
\end{equation}
This variable transformation leaves $\tilde{\mathcal{A}}$ invariant, but brings $\tilde{\mathcal{B}}$ to diagonal form, with unequal diagonal elements. The stochastic effects thus  correspond to \emph{anisotropic diffusion}. However, for large system size $N$, the effects of the anisotropy are  washed out: The system's  motion on the deterministic cycles, described by $\mathcal{A}$, occurs on a much faster timescale then the anisotropic diffusion, resulting in an averaging over the different directions.

To calculate the time evolution of the average deviation from the reactive fixed point, $\partial_t\langle\mathcal{R}^2(t) \rangle$, we start from the
the fluctuations in ${\bf y}$, which  fulfill the equations
\begin{align}
\partial_t\langle y_iy_j\rangle = \tilde{\mathcal{A}}_{ik}\langle y_ky_j \rangle+\tilde{\mathcal{A}}_{jk}\langle y_ky_i\rangle +\tilde{\mathcal{B}}_{ij}\quad.
\end{align}
Using Eq.~(\ref{R-y}), we obtain
\begin{equation}
\partial_t\langle\mathcal{R}^2(t) \rangle=\tilde{\mathcal{B}}_{AA}+\tilde{\mathcal{B}}_{BB}\quad .
\end{equation}
The dependence on the deterministic part has dropped out, and the solution to the above equation with initial condition $\langle\mathcal{R}^2(t=0) \rangle=0$ is a linear increase in the rescaled time $u\equiv t/N$:
\begin{equation}
\langle\mathcal{R}^2(t)    \rangle=\frac{3}{2}\frac{\omega_0^2}{k_A+k_B+k_C}\bigg[\frac{k_A^2+k_Ak_C+k_C^2}{k_A^2k_C^2}\omega_0^2 +1 \bigg]\frac{t}{N}\quad,
\end{equation}
valid around the reactive fixed point.
As in the case of equal reaction rates, we have a linear dependence $\langle\mathcal{R}^2(t) \rangle\sim t/N$, corresponding to a two-dimensional random walk.

As a conclusion of the above discussion, the general case of unequal reaction rates qualitatively reproduces the behavior of the before discussed simplest situation of equal rates (confirmed by stochastic simulations). The latter turns out to already provide a comprehensive understanding of the system.

\end{appendix}



\end{document}